\documentclass[11pt]{article}
\usepackage{epsfig}
\usepackage{rotating}

\newcommand{\BABARPubYear}    {08}

\newcommand{\BABARConfNumber} {018}
\newcommand{\SLACPubNumber} {13334}

\newcommand{\om}{\ifmmode {w} \else {$w$}\fi}
\newcommand{\Aone}{\ifmmode {{\cal A}_1} \else {${\cal A}_1$}\fi}
\newcommand{\rha}{\ifmmode{\mbox{\rho^2_{{\cal A}_1}}} \else {\mbox{$\rho^2_{{\cal A}_1}$}}\fi}
\newcommand{\rhv}{\ifmmode{\mbox{\rho^2_{{\cal V}_1}}} \else {\mbox{$\rho^2_{{\cal V}_1}$}}\fi}
\newcommand{\AoneVcb}{\ifmmode {{\cal A}_1(1)|V_{cb}|}\else {${\cal A}_1(1)|V_{cb}|$}\fi}

\input babarsym

\long\def\inst#1{\par\nobreak\kern 4pt\nobreak
    {\it #1}\par\vskip 10pt plus 3pt minus 3pt}

\begin{document}
{\pagestyle{empty}

\begin{flushright}
\babar-CONF-\BABARPubYear/\BABARConfNumber \\
SLAC-PUB-\SLACPubNumber \\
\end{flushright}

\par\vskip 4cm

\begin{center}
\Large \bf Measurement of {\boldmath  $|V_{cb}|$} and the form-factor slope for  {\boldmath  $\Bbar \to D\ell^-\bar{\nu}_{\ell}$} decays on the recoil of fully reconstructed {\boldmath  $B$} mesons
\end{center}
\bigskip

\begin{center}
\large The \babar\ Collaboration\\
\mbox{ }\\
\today
\end{center}
\bigskip \bigskip

\begin{center}
\large \bf Abstract
\end{center}
We present a measurement of the CKM matrix element $|V_{cb}|$ and the form-factor slope $\rho^2$ for  $\Bbar \to D \ell^- \bar{\nu}_{\ell}$ decays based on 
417 fb$^{-1}$ of data collected at the $\Upsilon(4S)$ resonance with the \babar\
detector. The semileptonic decays are selected in \BB\ events in which the
hadronic decay of the second $B$ meson is fully reconstructed. From the measured differential decay rate of the signal decay
we determine
${\cal G}(1) |V_{cb}|= (43.0 \pm 1.9 \pm 1.4)\times 10^{-3}$, $\rho^2 = 1.20 \pm 0.09 \pm 0.04$, where ${\cal G}(1)$ is the  hadronic form factor at the point of zero recoil. Using a lattice calculation for ${\cal G}(1)$ we extract $|V_{cb}|= (39.8\pm 1.8 \pm 1.3 \pm 0.9) \times 10^{-3}$, where the stated errors refer to the statistical, systematic, and form factor uncertainties.  We also present a measurement of the exclusive branching fractions, 
${\cal B}(\B^- \to D^0 \ell^- \bar{\nu}_{\ell}) = (2.31 \pm 0.08 \pm 0.07)\%$
and ${\cal B} (\Bzb \to D^+ \ell^- \bar{\nu}_{\ell})=(2.23 \pm 0.11 \pm 0.08)\%$.

\vfill
\begin{center}

Submitted to the 34$^{\rm th}$ International Conference on High-Energy Physics, ICHEP 08,\\
29 July---5 August 2008, Philadelphia, Pennsylvania.

\end{center}

\vspace{1.0cm}
\begin{center}
{\em Stanford Linear Accelerator Center, Stanford University, 
Stanford, CA 94309} \\ \vspace{0.1cm}\hrule\vspace{0.1cm}
Work supported in part by Department of Energy contract DE-AC02-76SF00515.
\end{center}

\newpage
} 

\begin{center}
\small

The \babar\ Collaboration,
\bigskip

%
B.~Aubert,
M.~Bona,
Y.~Karyotakis,
J.~P.~Lees,
V.~Poireau,
E.~Prencipe,
X.~Prudent,
V.~Tisserand
\inst{Laboratoire de Physique des Particules, IN2P3/CNRS et Universit\'e de Savoie, F-74941 Annecy-Le-Vieux, France }
J.~Garra~Tico,
E.~Grauges
\inst{Universitat de Barcelona, Facultat de Fisica, Departament ECM, E-08028 Barcelona, Spain }
% ...

\end{center}\newpage

\section{Introduction}
\label{sec:Introduction}

The study of the semileptonic decay $\Bbar \to D \ell^- \bar{\nu}_{\ell}$ 
is interesting in many aspects. In the standard model,
the rate of this weak decay is proportional to the square
of the Cabibbo-Kobayashi-Maskawa (CKM)~\cite{CKM} matrix element
$|V_{cb}|$, which is a measure of the weak coupling of the
$b$ to the $c$ quark. 

The decay rate is also proportional to the square of the hadronic matrix element,  
which accounts for the effects of strong
interactions in the $\Bbar \to D$ transition. In the limit of very small masses of the lepton,
$\ell=e$ or $\mu$, their effect can be 
parameterized by a single form factor ${\cal G} (\om)$, where 
the variable $\om$ (see below) is linearly related to the 
momentum transfer squared $q^2$ of the $B$ meson to the $D$ meson. 

The extraction of  $|V_{cb}|$ relies on the measurement of
differential decay rates of semileptonic $B$ decays. 
The precise determination requires corrections to
the prediction for the normalization 
at $\om=1$ in the 
context of Heavy Quark Symmetry~\cite{HQS}, as well as a
measurement of the variation of the form factors near
the kinematic limit, $\om=1$, where the decay rate goes to zero as
the phase space vanishes.

Measurements of $|V_{cb}|$  based on
studies of the differential decay rate for $\Bbar\to D\ell^-\bar{\nu}_{\ell}$ decays  have previously been reported by Belle~\cite{BELLE_d}, CLEO~\cite{CLEO_d} and ALEPH~\cite{ALEPH_d}. 

In this paper, we present preliminary measurements of the differential decay rates, separately for $\Bzb \to D^+ \ell^- \bar{\nu}_{\ell}$
and $B^- \to D^0 \ell^- \bar{\nu}_{\ell}$ ~\cite{charge}. Semileptonic decays are selected in \BB\ events in which an hadronic decay of the second $B$ meson is also fully reconstructed. This leads to a very clean sample of events and also provides a precise measurement for the variable $q^2$,  and thereby $\om$.  
The measurement of the total branching fractions and the extraction of $|V_{cb}|$ requires an absolute normalization.
We use a sample of inclusive semileptonic decays, $\Bbar \to X \ell^- \bar{\nu}_{\ell}$, where only the charged lepton is reconstructed.  This choice
reduces the uncertainty of the normalization, because the lepton is also selected from a sample tagged by hadronic decays of the second $B$ meson.

\section{Parameterization of the Decay Rate}

In the limit of small lepton masses the partial decay rate for $\Bbar\to D\ell^-\bar{\nu}_{\ell}$ can be expressed in terms of a single form factor, ${\cal G}(\om)$, 
\begin{eqnarray}
\label{eq:diffrate_dlnu}                                         
 \frac{{\rm d}\Gamma(B\to D\ell\nu)}{{\rm d}\om}  ~ &=& ~ \frac{G^2_F}{48 \pi^3 \hbar} M^3_{D} (M_{B}+M_{D})^2 
                            ~({\om^2-1})^{3/2}~  \mid V_{cb} \mid^2 ~ {\cal G}^2 (\om),       
\end{eqnarray}

\noindent where $G_F$ is the Fermi coupling constant, and $M_{B}$ and $M_{D}$ are the masses of the $B$ 
and $D$ mesons. The variable $\om$ denotes the product of the $B$ and $D$ meson four-velocities $V_B$ and $V_D$, 

\begin{equation}
\nonumber   \om = V_B\cdot V_D=\frac{(M_{B}^2 + M_{D}^2 - q^2)}{(2M_{B} M_{D})},
\label{eq:w}
\end{equation}

\noindent where $q^2 \equiv (p_{B}-p_{D})^2$, and $p_B$ and $p_D$ refer to the four-momenta of the $B$ and $D$ mesons.  Its lower limit,
$\om=1$, corresponds to zero recoil of the $D$ meson, $i.e.$ the maximum $q^2$.  The upper limit, $\om = 1.59$,  corresponds to $q^2=0$ and the maximum $D$ momentum.   Since the $B$ momentum is known from the fully reconstructed $B_\mathrm{tag}$ in the same event, $\om$ can be reconstructed with good precision, namely to 
$\sim 0.01$, which corresponds to about $2\%$ of the full kinematic range.

In the limit of infinite quark masses ${\cal G}(\om)$ coincides with the 
Isgur-Wise function~\cite{IW}. This function is
normalized to unity at zero recoil. Corrections to this prediction have recently been calculated with improved precision, based on unquenched lattice QCD~\cite{Okamoto}, specifically
${\cal G}(1) = 1.074\pm 0.018\pm 0.016$.
Thus $|V_{cb}|$ can be extracted by extrapolating the differential
decay rates to $\om = 1$. To reduce the uncertainties associated with this extrapolation, constraints on the shape of the form factors are highly 
desirable. Several functional forms have been proposed~\cite{Neuold}. We adopt the parameterization based on analyticity and positivity of the QCD functions which describe the local currents~\cite{Neunew}.  Specifically,
${\cal G}(\om)$ is expressed as a polynomial in $z$, 
\begin{equation}
\nonumber  {\cal G}(\om) ~=~ {\cal G}(1) \left[ 1-8\rho^2 z +(51\rho^2 -10)z^2-
      (252\rho^2 -84)z^3 \right],
\label{eq:g_dlnu}
\end{equation}

\noindent where the variable $z$ is defined as $z=(\sqrt{\om+1}-\sqrt{2})/(\sqrt{\om+1}+\sqrt{2})$.
This formulation expresses the non-linear dependence of the form factor on $\om$ in terms of a single shape parameter, $\rho^2$.


\section{The \babar\ Detector and Dataset}
\label{sec:babar}
This analysis is based on 417 fb$^{-1}$ of data collected at the $\Upsilon(4S)$ resonance with the \babar\ detector at the \pep2\ storage rings. The corresponding number of produced \BB\ pairs is 460 million. In addition, 
40 fb$^{-1}$  of data, recorded at a center-of-mass energy 40 MeV below the $\Upsilon(4S)$ resonance, are used to study background from $e^+e^- \to f\bar{f}~(f=u,d,s,c,\tau)$ events (continuum production). The \babar\ detector is described in detail elsewhere~\cite{detector}. Charged-particle trajectories are measured by a 5-layer double-sided silicon vertex tracker and a 40-layer drift chamber, both operating in a 1.5-T magnetic field. Charged-particle identification is provided by the average energy loss (d$E$/d$x$) in the tracking devices and by an internally reflecting ring-imaging Cherenkov detector. Photons are detected by a CsI(Tl) electromagnetic calorimeter (EMC). Muons are identified by the instrumented magnetic-flux return. 
A detailed GEANT4-based Monte Carlo (MC) simulation~\cite{Geant} of \BB\ and continuum events has been used to study the detector response, its acceptance, and to test the analysis procedure. The simulation models the signal $\Bbar \to D \ell^- \bar{\nu}_{\ell}$ decays using the ISGW2 model~\cite{ISGW}, and these are then reweighted to the HQET model described above.
Other semileptonic decays that contribute to background are simulated as follows: for $\Bbar \to D^* \ell^- \bar{\nu}_{\ell}$
decays we use form factor parametrizations based on Heavy Quark Effective Theory (HQET) with parameters determined by the \babar\ collaboration~\cite{BABARFF}, for $\Bbar \to D^{**}(\rightarrow D^{(*)} \pi) \ell^- \bar{\nu}_{\ell}$ decays we use the ISGW2 model~\cite{ISGW}, and for decay involving non-resonant charm states, $\Bbar \to D^{(*)} \pi \ell^- \bar{\nu}_{\ell}$,  we adopt the prescription of the Goity-Roberts model~\cite{Goity}.  
The MC simulation includes radiative effects such as bremsstrahlung in the detector material. 
QED final state radiation is modeled by PHOTOS~\cite{photos}, and decays with radiative photons are included in the signal sample.

\section{Event Selection}
\label{sec:Analysis}
We select semileptonic $B$ meson decays in events containing
a fully reconstructed $B$ meson ($B_\mathrm{tag}$), which allows us to constrain the kinematics, to reduce the combinatorial background, and to determine the charge and flavor of the signal $B$.

The analysis exploits the presence of two charmed mesons in the final state:
one is used for the exclusive reconstruction of the $B_\mathrm{tag}$, and the other one
for the reconstruction of the semileptonic $B$ decay.

We first reconstruct the semileptonic $B$ decay, selecting a lepton with momentum in the  center-of-mass (CM) frame $p^*_{\ell}$  higher than 0.6 GeV/$c$. Electrons from photon conversion and $\pi^0$ Dalitz decays are removed using a dedicated algorithm, which performs the reconstruction of vertices between tracks of opposite charge whose invariant mass is compatible with a photon conversion or a $\pi^0$ Dalitz decay.  Candidate $D^0$ mesons, with the correct 
charge correlation with the lepton, are reconstructed
in the $K^-\pi^+$, $K^- \pi^+ \pi^0$, $K^- \pi^+ \pi^+ \pi^-$,
$K^0_S \pi^+ \pi^-$, $K^0_S \pi^+ \pi^- \pi^0$, $K^0_S \pi^0$, $K^+ K^-$,
$\pi^+ \pi^-$, and $K^0_S K^0_S$ channels, and $D^+$ mesons in the
$K^- \pi^+ \pi^+$, $K^- \pi^+ \pi^+ \pi^0$, $K^0_S \pi^+$, $K^0_S \pi^+ \pi^0$,
$K^+ K^- \pi^+$, $K^0_S K^+$, and $K^0_S \pi^+ \pi^+ \pi^-$ channels.
$D$ candidates are selected within 2$\sigma$ of the $D$ mass, with $\sigma$ typically around 8 MeV$/c^{2}$. 
In events with multiple $\Bbar \to D \ell^- \bar{\nu}_{\ell}$ candidates, the candidate with the largest $D$-$\ell^-$ vertex fit probability is selected.

We reconstruct $B_\mathrm{tag}$ decays of the type $\Bbar \rightarrow D^{(*)} Y$, where 
$Y$ represents a collection of hadrons with a total charge of $\pm 1$, composed
of $n_1\pi^{\pm}+n_2 K^{\pm}+n_3 K^0_S+n_4\pi^0$, where $n_1+n_2 \leq  5$, $n_3
\leq 2$, and $n_4 \leq 2$. Using $D^0(D^+)$ and $D^{*0}(D^{*+})$ as seeds for $B^-(\Bzb)$ decays, we reconstruct about 1000 decay modes.

The kinematic consistency of a $B_\mathrm{tag}$ candidate with a $B$ meson is checked using two variables: the beam-energy
substituted mass $m_{ES}=\sqrt{s/4-\vec{p}_B^{~2}}$, and the energy difference $\Delta E = E_B -\sqrt{s}/2$. Here $\sqrt{s}$ refers to the total CM  energy, and $\vec{p}_B$ and $E_B$ denote the momentum and energy of the $B_\mathrm{tag}$ candidate in the CM frame. For correctly identified $B_\mathrm{tag}$ decays, the $m_{ES}$ distribution peaks at the $B$ meson mass, while $\Delta E$ is consistent
with zero.
We select a $B_\mathrm{tag}$ candidate in the signal region
defined as 5.27~GeV/$c^2$ $< m_{ES} <$ 5.29~GeV/$c^2$, excluding $B_\mathrm{tag}$ candidates with
 daughter particles in common with the charm meson or
with the lepton from the semileptonic $B$ decay. 
In the case of multiple $B_\mathrm{tag}$ candidates, we select the one with the smallest
$|\Delta E|$ value. For charged $B$ events, the $B_\mathrm{tag}$ and the $D\ell^-$ candidates are required to have the correct charge-flavor correlation. We do not apply any correction to account for $B^0-\Bzb$ mixing effects, because they are found to be negligible. 
Cross-feed effects, $i.e.$ $B^-_\mathrm{tag} (\Bzb_\mathrm{tag})$ candidates erroneously reconstructed as a neutral~(charged) $B$,  are subtracted using the  MC simulation.

Semileptonic $B$ decays are identified by the missing mass squared, defined as:

\begin{equation}
 m^2_\mathrm{miss} = \left[ p(\Upsilon(4S)) -p(B_\mathrm{tag}) - p(D) - p(\ell)\right]^2
\end{equation}

\noindent in terms of the particle
four-momenta. For correctly reconstructed signal events, the only missing particle is the neutrino, and the $m^2_\mathrm{miss}$ peaks at zero. Other semileptonic $B$ decays, like  $\Bbar \to D^* \ell^- \bar{\nu}_{\ell}$ and 
$\Bbar \to D^{**}\ell^- \bar{\nu}_{\ell}$, where one particle is not reconstructed (feed-down), spread to higher values of $m^2_\mathrm{miss}$. The use of the full-reconstruction technique  results in a $m^2_\mathrm{miss}$ resolution of 0.04~GeV$^2$/$c^4$, an order of magnitude better than achieved in
non-tagged analyses~\cite{exampArgus}.

\section{Measurement of $|V_{cb}|$ and $\rho^2$}
\label{sec:fit}

We measure $|V_{cb}|$ and the form-factor slope $\rho^2$ by a fit to the $\om$ distribution for 
$\Bbar\to D\ell^-\bar{\nu}_{\ell}$ decays.    

Data and Monte Carlo events are  collected in ten equal-size $\om$ bins. 
The few events that have $w <1.0$ or $w>1.59$ due to resolution effects, are collected in the first bin and the last bin, respectively. 

To obtain the semileptonic $\Bbar \to D\ell^-\bar{\nu}_{\ell}$ signal yield in the different $\om$ intervals, 
we perform a one-dimensional extended binned maximum likelihood fit to the $m^2_\mathrm{miss}$ distributions, 
based on a method developed by R.~Barlow and C.~Beeston~\cite{Barlow}. The fitted data samples are assumed 
to contain four different types of events:

\begin{itemize}
\item signal $\Bbar \to D \ell^- \bar{\nu}_{\ell}$, 
\item feed-down semileptonic $B$ decays,
\item \BB\ and continuum background,
\item fake lepton events. 
\end{itemize}

\noindent We use the Monte Carlo predictions for the different semileptonic $B$ decay $m^2_\mathrm{miss}$ distributions to obtain the Probability Density Functions (PDFs) to fit the data distributions. The feed-down contributions are allowed to float independently in the different $w$ intervals. 
We use the off-peak data to provide the continuum background normalization, while the \BB\ normalization is taken from the MC. The shape of the continuum background predicted by the MC simulation is consistent with the one obtained from the off-peak data. 

\begin{figure}[!ht]
\centering
\epsfig{figure=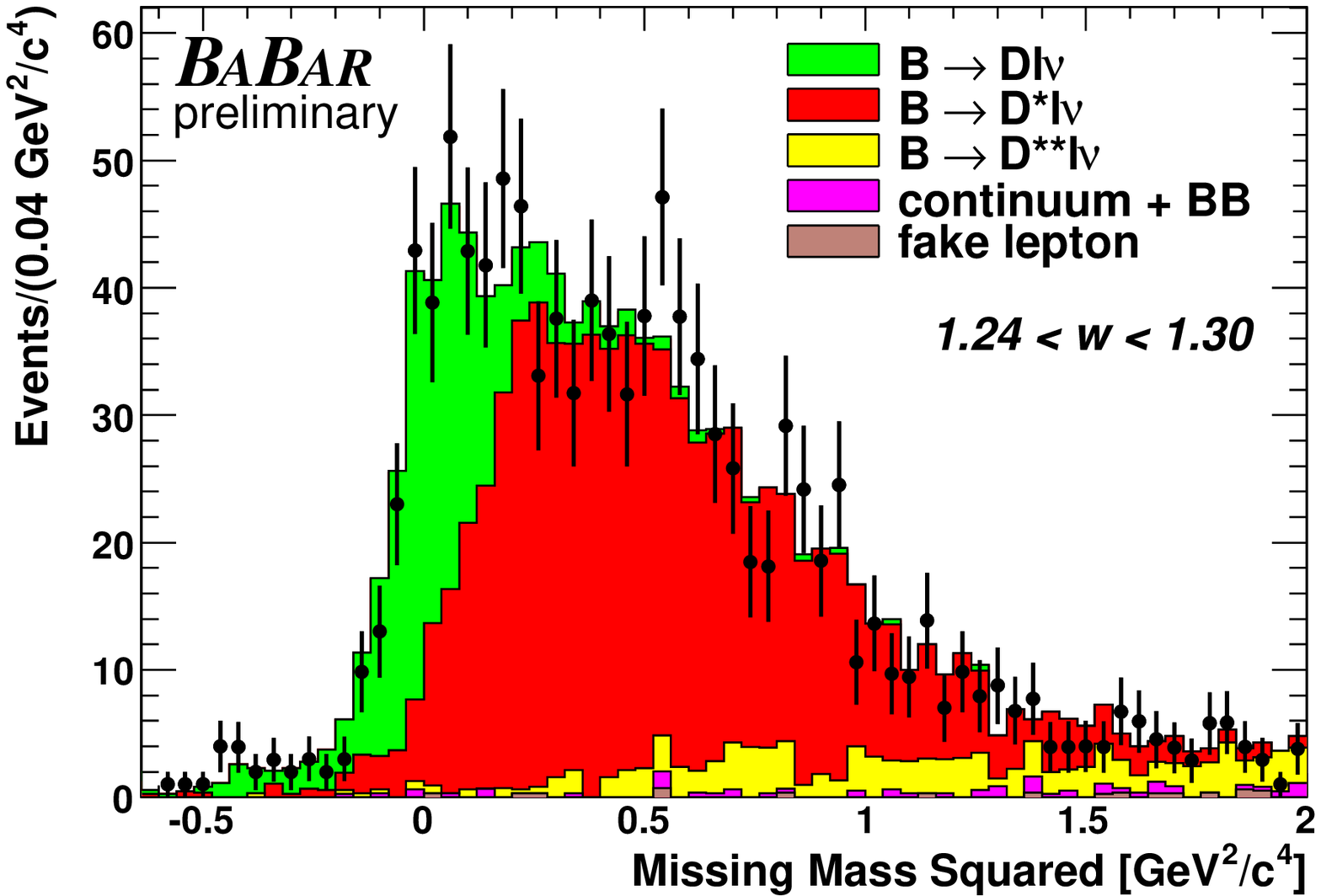,width=8.1cm}
\epsfig{figure=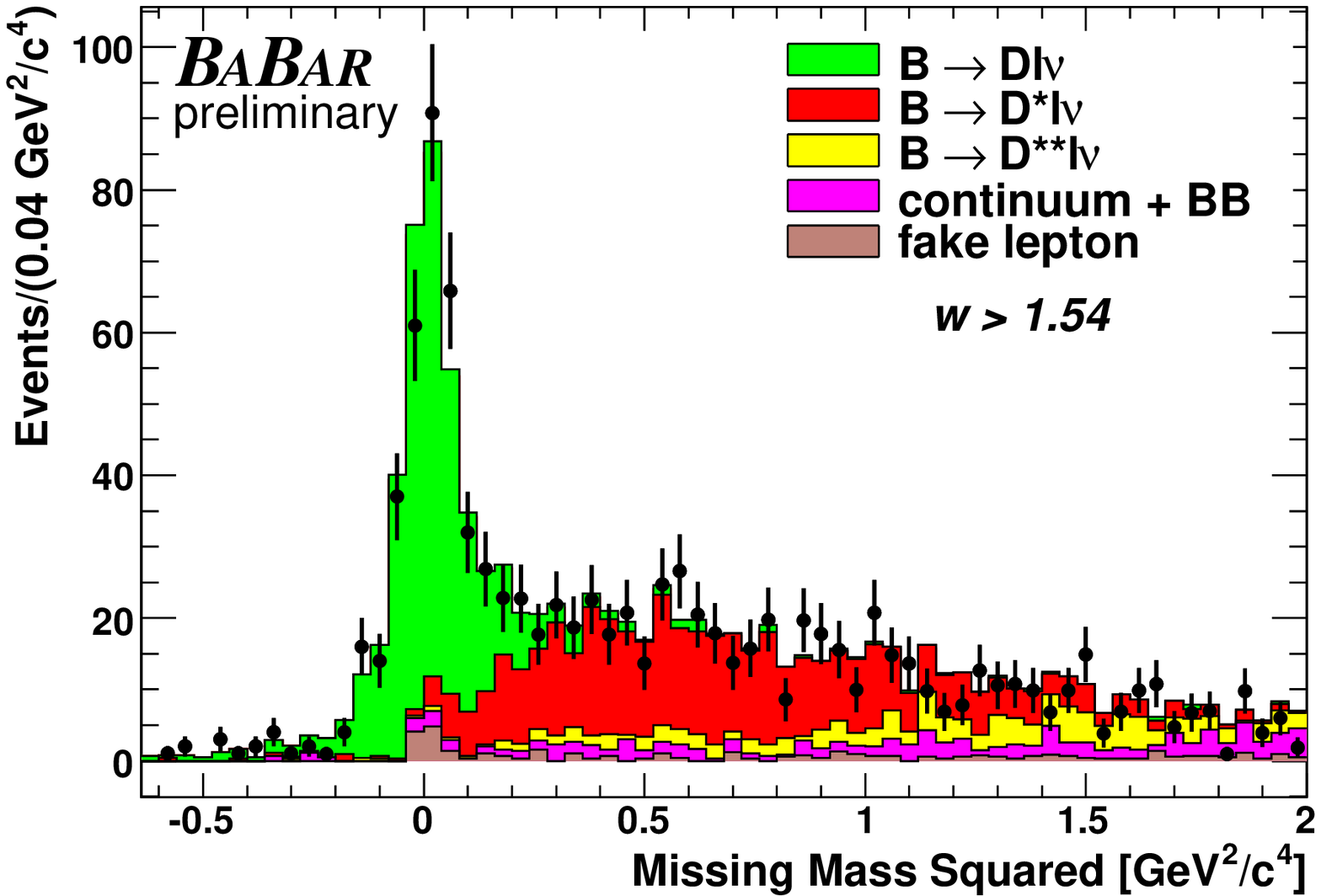,width=8.1cm}
\epsfig{figure=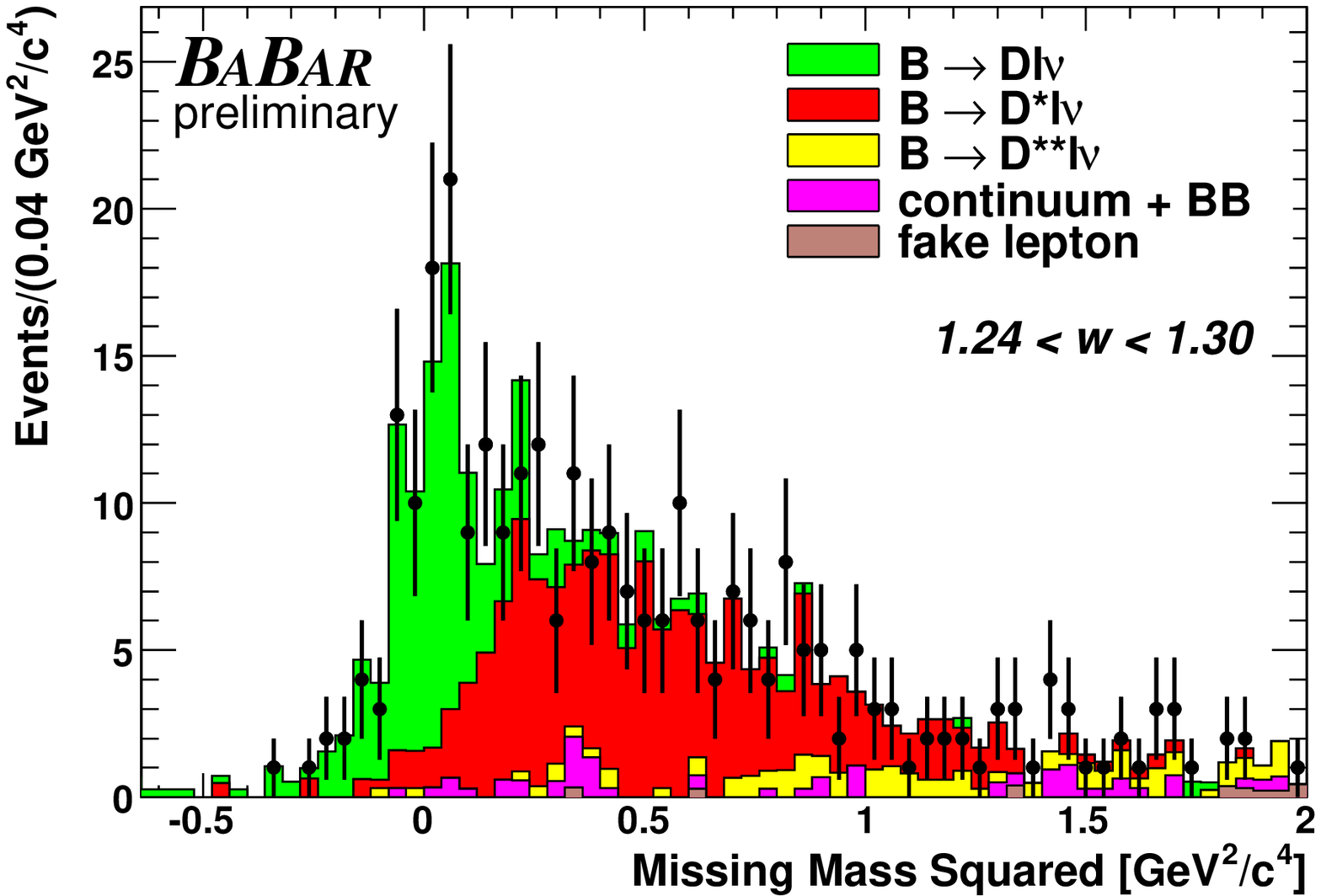,width=8.1cm}
\epsfig{figure=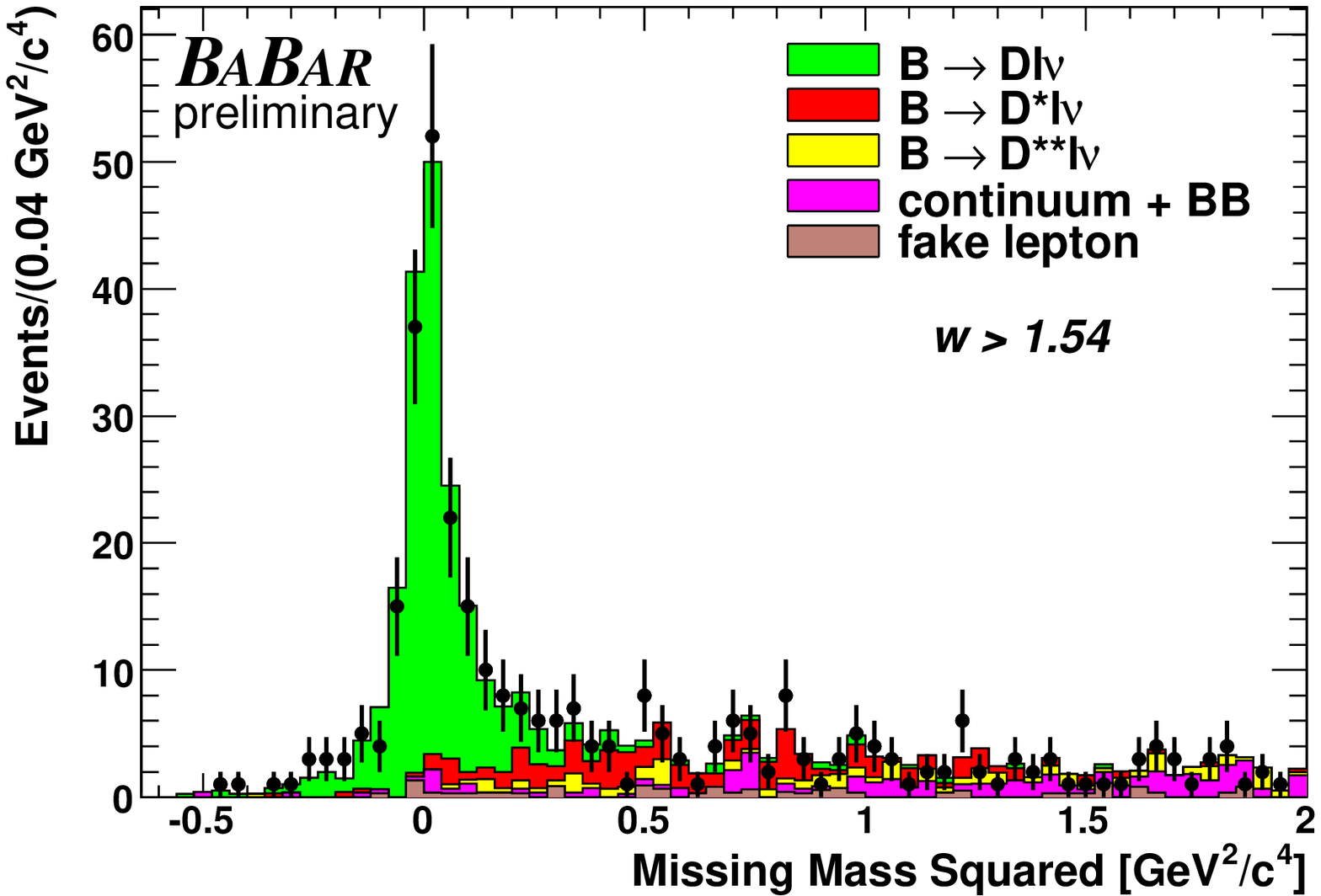,width=8.1cm}
\caption{Fit to the $m^2_\mathrm{miss}$ distribution, in two different $w$ intervals, 
for $B^- \to D^0 \ell^- \bar{\nu}_{\ell}$ (top row),  $\Bzb \to D^+ \ell^- \bar{\nu}_{\ell}$ (bottom row): 
the data (points with error bars) are compared to the results of the overall
 fit (sum of the solid histograms). The PDFs for the different fit components 
are stacked and shown in different colors.}
\label{fig:Fit1}
\end{figure}

The $m^2_\mathrm{miss}$ distributions for two different $w$ intervals are compared with the results of the 
fits in Fig.~\ref{fig:Fit1}.

\begin{figure}[!t]
\begin{center}
\begin{tabular}{ c c }
\includegraphics[width=8.2cm]{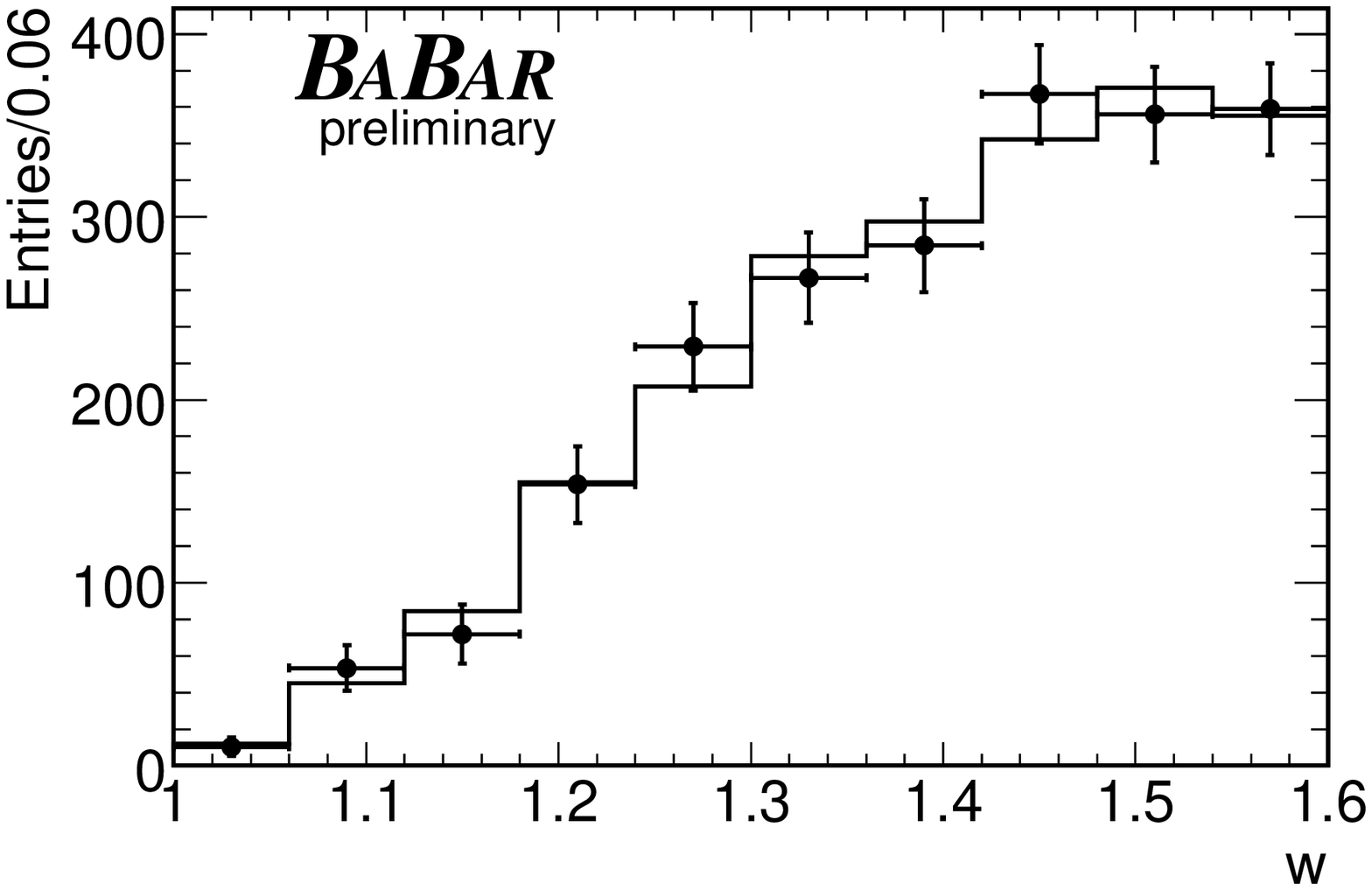}
 & \includegraphics[width=8.2cm]{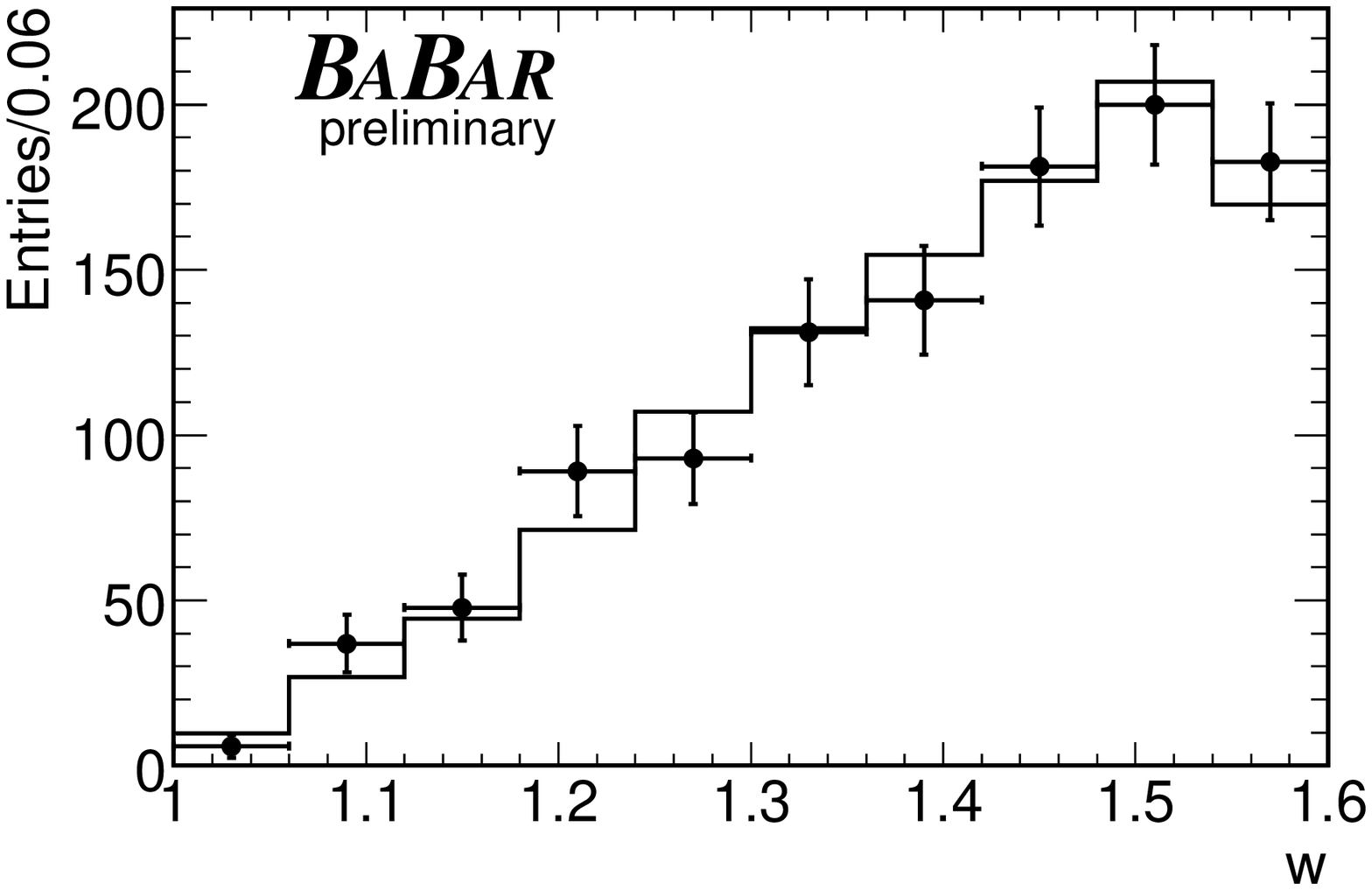}\\
\end{tabular}
\caption{\label{fig:dndw_plots} The $w$ distribution 
for $\Bbar \to D\ell^-\bar{\nu}_{\ell}$ events. Left: $B^- \to D^0\ell^-\bar{\nu}_{\ell}$, right: $\Bzb \to D^+\ell^-\bar{\nu}_{\ell}$. The data (points with error bars) are compared to the results of the overall
 fit (solid histogram). }
\end{center}\end{figure}

The signal yields in bins of $\om$ are displayed in Fig.~\ref{fig:dndw_plots}. 
A $\chi^2$ fit is performed on the $w$ distributions comparing the number of events observed in each 
interval to the MC signal prediction. 
The signal prediction is obtained by properly weighting each MC event. The $\chi^2$ is defined as:

\begin{equation}
\label{eq:chi2}
\chi^2 = \sum_{i=1}^{10} \frac{(N^i_{\rm data} - \sum_{j=1}^{N^i_{\rm MC}} W^i_j )^2}{ (\sigma^i_{\rm data})^2 +  \sum_{j=1}^{N^i_{\rm MC}} W{^i_j}^2},
\end{equation}

\noindent 

\noindent where the index $i$ runs over the ten $\om$ bins; $N^i_{\rm data}$ is the observed 
number of signal events found in the $i^{th}$ bin and the $\sigma^i_{\rm data}$ the corresponding error. 
The expected signal yields are calculated
at each step of the minimization from the reweighted sum of $N^i_{\rm MC}$ simulated events. 
The Monte Carlo is corrected event by event for all known differences in tracking, cluster reconstruction and particle 
identification determined from data control samples. 
The weight for each Monte Carlo event is computed as the product of two terms
 $W^i_j = W^{\cal L} \times W^{i,\rm theo}_j$ where 

\begin{itemize}
\item $W^{\cal L}$ is an overall fixed scale factor that accounts for the relative luminosity of 
data and signal Monte Carlo events; this is obtained, as described below,  by normalizing to the inclusive yield of $\Bbar \to X\ell^-\bar{\nu}_{\ell}$ events, corrected for its reconstruction efficiency, and scaled to the absolute $\Bbar \to X\ell^-\bar{\nu}_{\ell}$ branching fraction~\cite{pdg}.
\item $ W_j^{i,\rm theo} = f^{\rm theo}(w_j^i; \rho^2, {\cal G}(1)\Vcb ) / f_{\rm MC}(w_j^i; \rho^2_{\rm MC}, {\cal G}(1)\Vcb_{\rm MC} )$ 
is the term that describes the differential decay rate. It depends on ${\cal G}(1) |V_{cb}|$ and $\rho^2$,  
and varies at each step of the minimization.
\end{itemize}

\noindent Here $w_{j}^{i}$ is the true $w$ of the MC event $j$, 
reconstructed in the $i$-bin. The function $f^{\rm theo}$ corresponds to the expressions in Eq.~\ref{eq:diffrate_dlnu} 
and~\ref{eq:g_dlnu}, while $f_{\rm MC}$ is the function used to generate $\Bbar \to D \ell^- \bar{\nu}_{\ell}$  
Monte Carlo events~\footnote{Corresponding to the CLN parameterization~\cite{Neunew}, with $\rho^2=1.17$.}.

We first fit separately the $\om$ distributions for the charged and neutral $\Bbar\to D\ell^-\bar{\nu}_{\ell}$ 
samples; we also perform a combined fit for $\Bbar\to D\ell^-\bar{\nu}_{\ell}$,  
assuming that $f_{00}+f_{+-}=1$, $i.e.$ that $B^+B^-$ and $B^0 \Bzb$ together saturate the $\Upsilon$(4S) decays. 
The value of the branching ratio is then computed by integrating the 
differential expression in Eq.~\ref{eq:diffrate_dlnu}.

\begin{figure}[!t]
\begin{center}
\begin{tabular}{ c c }
\includegraphics[width=8.2cm]{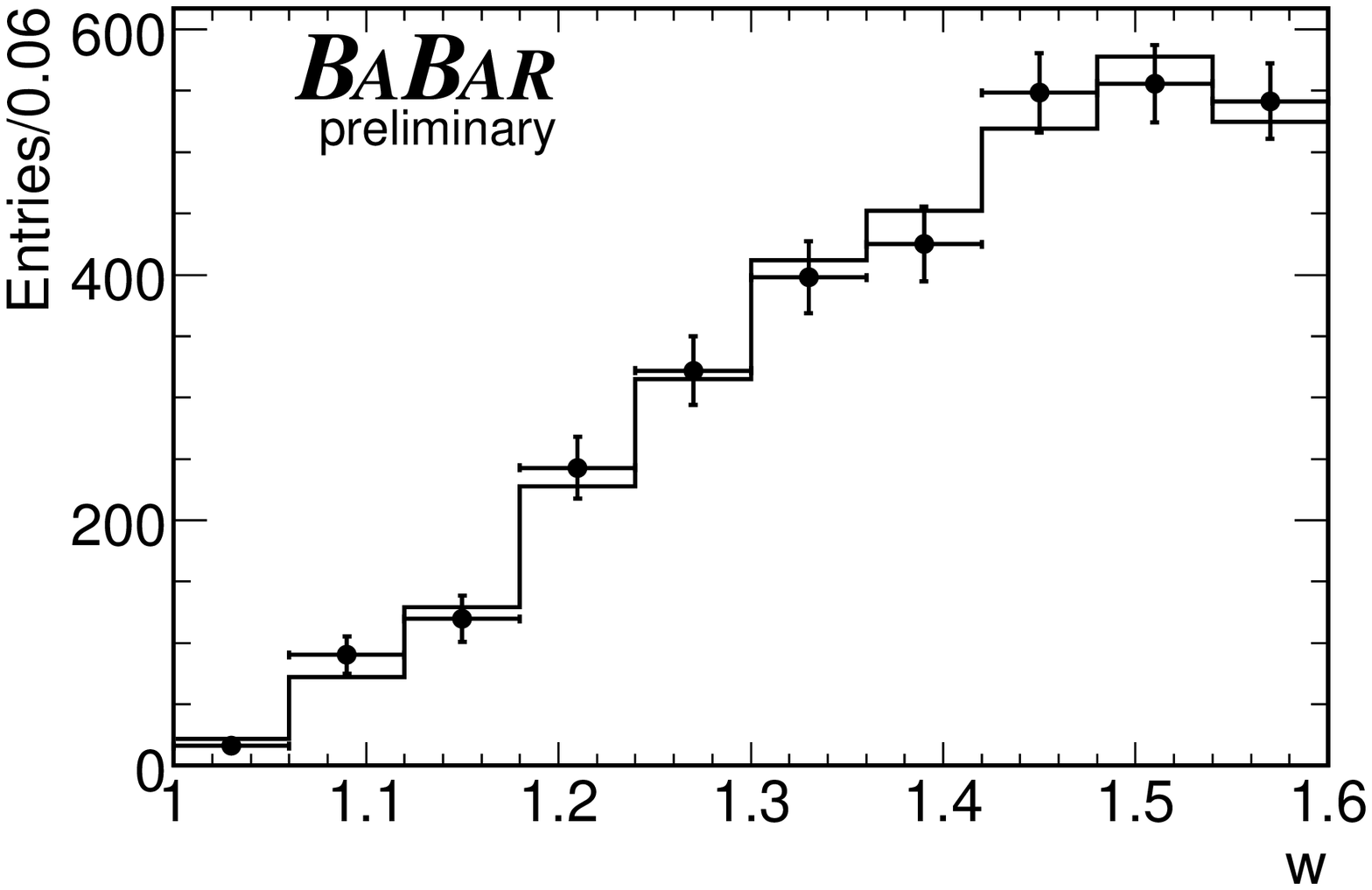} &
\includegraphics[width=8.2cm]{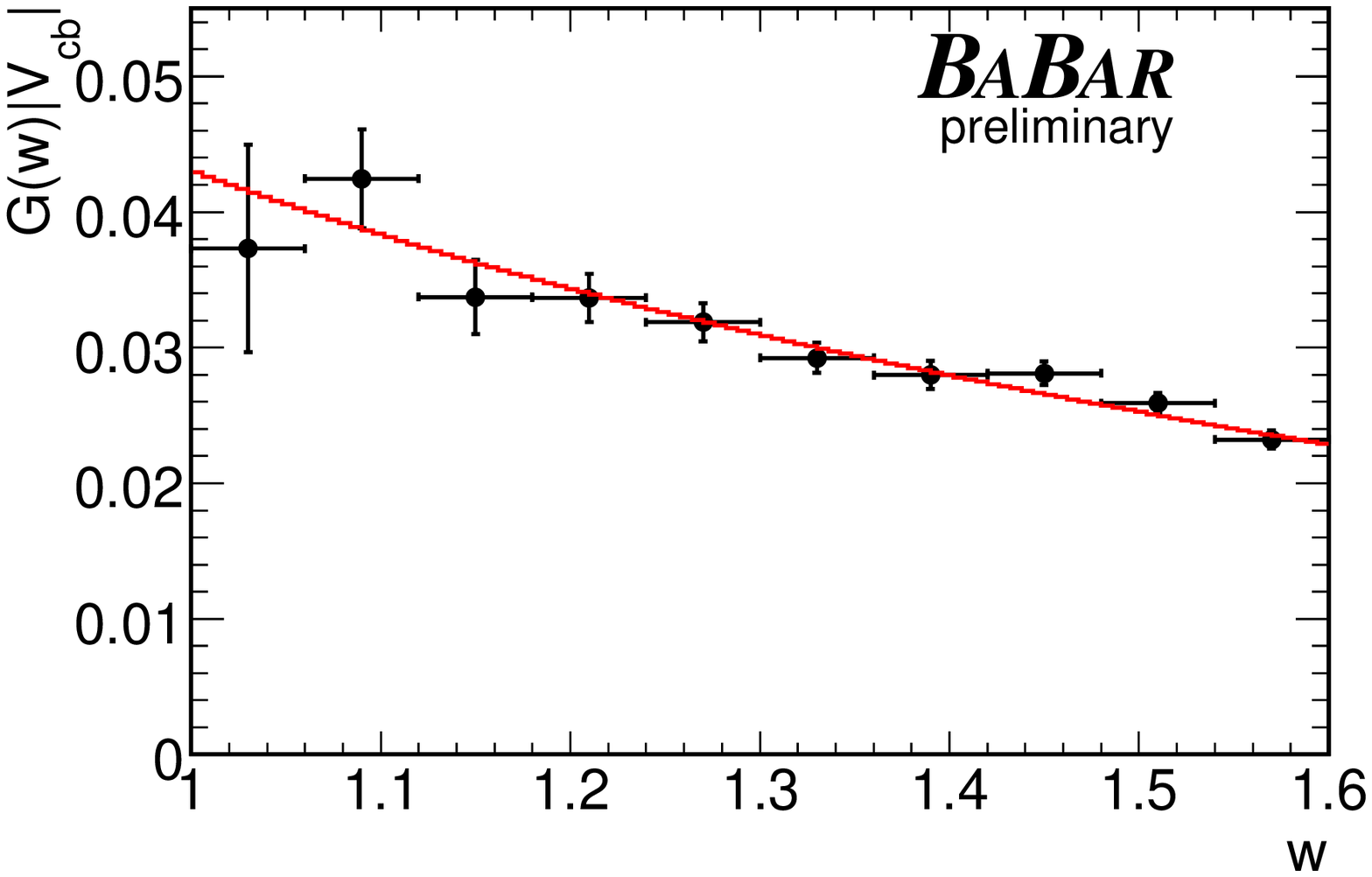}\\
\end{tabular}
\caption{\label{f:dndw-all} Left: $\om$ distribution obtained summing
together $B^- \to D^0\ell^-\bar{\nu}_{\ell}$ and $\Bzb \to D^+\ell^-\bar{\nu}_{\ell}$ yields. The data (points with error bars) are compared to the results of the overall
 fit (solid histogram). Right: ${\cal G}(w)|V_{cb}|$ distribution unfolded for the 
reconstruction efficiency, with the fit result superimposed. 
These plots are not corrected for the smearing in $w$.}
\end{center}\end{figure}

\begin{figure}[!t]
\begin{center}
\begin{tabular}{ c }
\includegraphics[width=8.2cm]{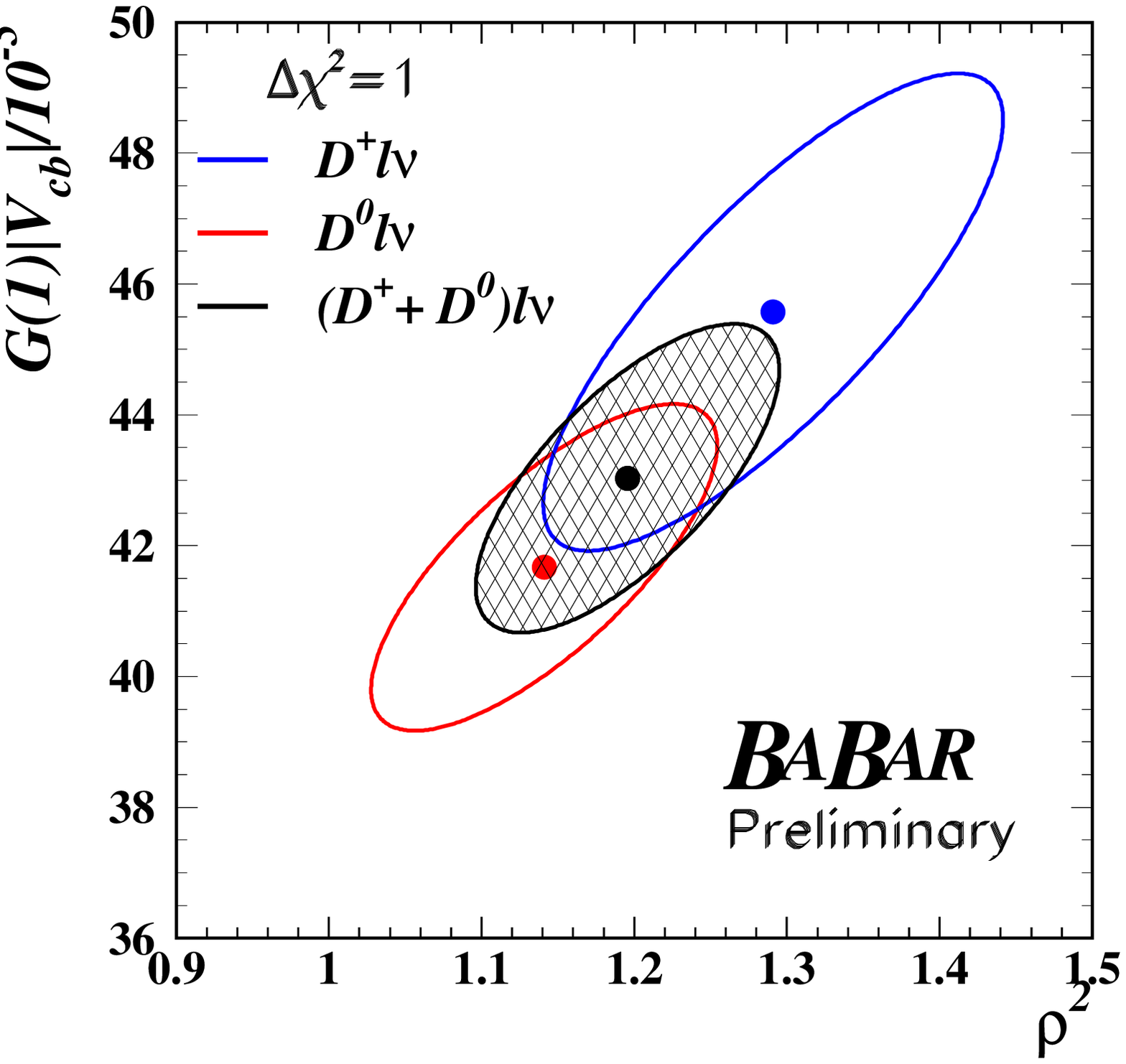}\\
\end{tabular}
\caption{\label{f:ellipses-Ds}(left) $\Delta \chi^2=1$ ellipses in the ${\cal G}(1)|V_{cb}|$ 
versus $\rho^2$ plane for the
$B^- \to D^0\ell^-\bar{\nu}_{\ell}$ decay, $\Bzb \to D^+\ell^-\bar{\nu}_{\ell}$ decay and the $\Bbar \to D\ell^-\bar{\nu}_{\ell}$ combined fit (ellipse in red). The systematic uncertainties are taken into account. }
\end{center}\end{figure}

In Fig. \ref{fig:dndw_plots} we show the comparison between the data and  the fit results separately for $B^- \to D^0\ell^-\bar{\nu}_{\ell}$ and $\Bzb \to D^+ \ell^- \bar{\nu}_{\ell}$ .  The corresponding distributions for the combined fit are shown in Fig. \ref{f:dndw-all}.
The measured values of ${\cal G}(1)|V_{cb}|$ and $\rho^2$, with the corresponding correlation $\rho_\mathrm{corr}$ obtained from the fit,  are reported in Table \ref{t:datafit} and shown in Fig.~\ref{f:ellipses-Ds}.

\begin{table}[!]
\centering
\caption{\label{t:datafit} Fit results for each sample. In the last column we report the results for the 
$\Bzb$ and $B^-$ combined fit, where the branching fraction refers to $\Bzb$ decays. We also report the signal yields and the reconstruction efficiencies, integrated over the full $w$ range. Only the statistical errors are reported here.}
\begin{tabular}{l|cc|c}
\hline
\hline
	           	 & $B^- \to D^0\ell^-\bar{\nu}_{\ell}$            & $\Bzb \to D^+\ell^-\bar{\nu}_{\ell}$		 & $\Bbar \to D\ell^-\bar{\nu}_{\ell}$       \\
\hline 
${\cal G}(1)\Vcb\cdot 10^3$ &41.7$\pm$  2.1	   &45.6$\pm$ 3.3	& 43.0$\pm$ 1.9	\\
$\rho^2$            	    & 1.14$\pm$  0.11	   & 1.29$\pm$ 0.14	& 1.20$\pm$ 0.09\\
$\rho_\mathrm{corr}$               & 0.943                 & 0.950              &   0.952      \\
$\chi^2/ndf$             & 3.4/8		   & 5.6/8			& 9.9/18		  \\
\hline
Signal Yield & 2147 $\pm$ 69 &  1108 $\pm$ 45 & - \\
Recon. efficiency & $(1.99 \pm 0.02)\cdot 10^{-4}$  & $(1.09 \pm 0.02) \cdot 10^{-4}$& -\\
${\cal B}$		&(2.31$\pm$ 0.08)$\%$ & (2.23$\pm$ 0.11)$\%$	& (2.17$\pm$ 0.06)$\%$\\
\hline
\hline
\end{tabular}
\end{table}

\begin{figure}[!h]
\centering
\epsfig{figure=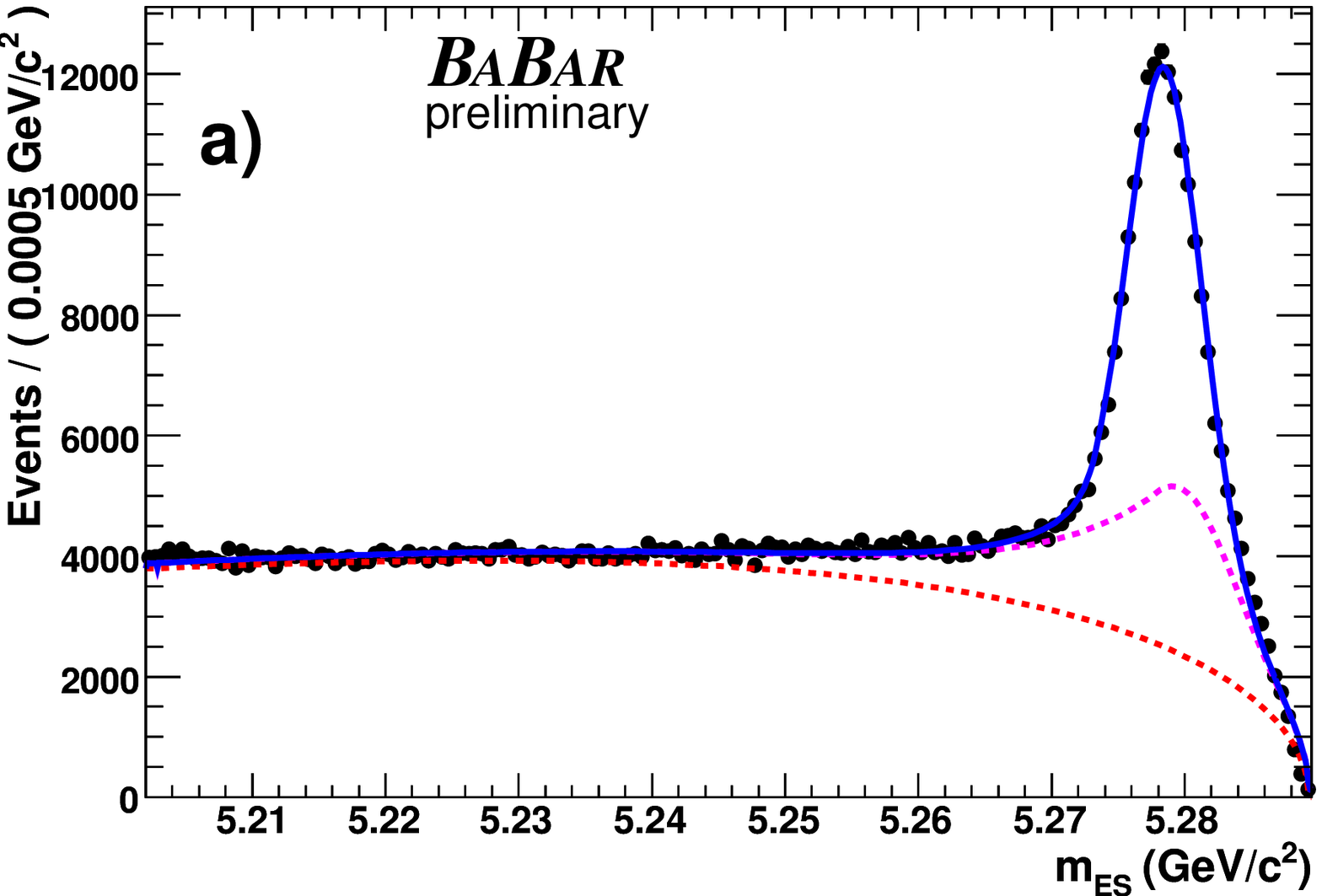,width=8.1cm}
\epsfig{figure=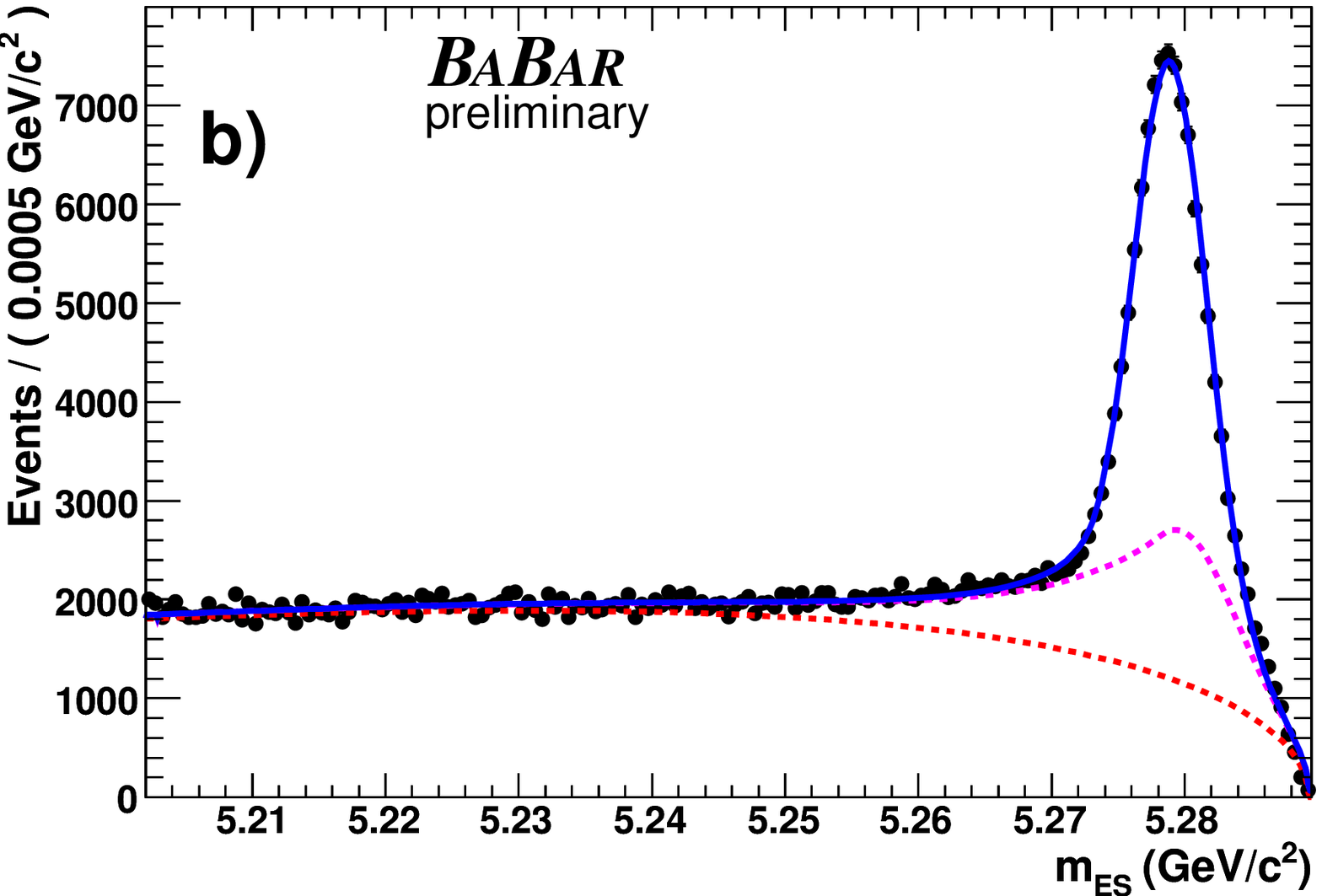,width=8.1cm}
\caption{$m_{ES}$ distributions of the a) $B^- \to X \ell^- \bar{\nu}_{\ell}$, and b) $\Bzb \to X \ell^- \bar
{\nu}_{\ell}$ samples. The data (points with error bars) are compared to the result of the fit (solid line). The dashed lines
 show the broad-peaking component and the sum of the combinatorial and continuum background.}
\label{fig:mesB}
\end{figure}

For the measurement of ${\cal G}(1)|V_{cb}|$, in order to reduce the systematic uncertainty, we use 
as normalization a sample of inclusive $\Bbar \to X \ell^- \bar{\nu}_{\ell}$ decays.
These events are selected by identifying a charged lepton with CM momentum greater than 0.6 GeV/$c$. In the case of multiple $B_\mathrm{tag}$ candidates, we select the one reconstructed in the decay channel with the highest purity, defined as the fraction of signal events in the $m_{ES}$ signal region. We require the lepton and the $B_\mathrm{tag}$ to have the correct charge  correlation and the lepton track to not be used to reconstruct the $B_\mathrm{tag}$ candidate. 
Background components peaking in the $m_{ES}$ signal region include cascade $B$ meson decays (i.e. the lepton does not come directly from the $B$) and hadronic decays where one of the hadrons is misidentified as a lepton. These backgrounds are subtracted by using the corresponding simulated Monte Carlo distributions.  The cascade-$B$ meson decays (17.6\% and 19.0\% of the total $m_{ES}$ distribution for charged and neutral $B$, respectively) are reweighted to account for differences between the branching fractions used in our Monte Carlo simulation and the latest experimental measurements~\cite{thorsten}.
The total yield for the inclusive $\Bbar \to X \ell^- \bar{\nu}_{\ell}$ decays is obtained from a maximum-likelihood fit to the $m_{ES}$ distribution of the $B_\mathrm{tag}$ candidates, using an ARGUS function~\cite{Argus} for the description of the combinatorial \BB\  and continuum background, and a Crystal Ball function~\cite{CrystallBall} for the signal. A broad peaking component is observed in the $m_{ES}$ signal region and is included in the signal definition. This is due to real $\Bbar \to X \ell^- \bar{\nu}_{\ell}$ decays for which, in the $B_\mathrm{tag}$ reconstruction, neutral particles are not reconstructed or are  interchanged with the semileptonic decays (e.g. a $\gamma$ from radiative $D^{*0}$ decay which belongs  to the $D^{*0}$ seed in the $B_\mathrm{tag}$ decay chain and is instead associated with a $B^- \to D^{*0} (D^0 \gamma) \ell^- \bar{\nu}_{\ell}$ decay).
This broad peaking component is modeled with additional Crystal Ball and ARGUS functions, whose parameters are fixed to the Monte Carlo prediction, except for the Crystal Ball mean value. 
Fig. \ref{fig:mesB} shows the $m_{ES}$ distribution for the $B_\mathrm{tag}$ candidates in the $B^- \to X \ell^- \bar{\nu}_{\ell}$ and $\Bzb \to X \ell^- \bar{\nu}_{\ell}$ samples. The fit yields 198,897 $\pm$ 1,578 signal events in the 
$B^- \to X \ell^- \bar{\nu}_{\ell}$ sample and 116,330 $\pm$ 1,088 signal events in the 
$\Bzb \to X \ell^- \bar{\nu}_{\ell}$ sample. The corresponding reconstruction efficiencies,
including the $B_\mathrm{tag}$ reconstruction, are 0.39\% and 0.25\%, respectively.

\section{Systematic Uncertainties}
\label{cha:sys}

Different sources of systematic uncertainties have been estimated and are given in Table~\ref{tab:Syst1}. We have grouped them into several categories.

Detector-related systematics may arise from
differences between the data and simulation of the track reconstruction and efficiency, particle
identification and neutral particle reconstruction.
The systematic uncertainty related to the reconstruction of charged tracks is determined by randomly removing a fraction of tracks corresponding to the uncertainty in the track finding efficiency, estimated on $e^+e^- \to \tau^+\tau^-$ data control samples.  The systematic uncertainty due to the reconstruction of neutral particles in the EMC is studied by varying the resolution and efficiency to match those found in data control samples.
We estimate the systematic uncertainty due to particle identification by varying the electron and muon identification efficiencies by 2\% and 3\%, respectively. The misidentification probabilities are varied by 15\% for both electrons and muons.

We evaluate the systematic uncertainties associated with the Monte Carlo simulation of the various signal and background processes. 
The uncertainty arising from radiative corrections is studied by comparing the results using PHOTOS with those obtained with PHOTOS turned off. We take 30\% of the difference as a conservative systematic uncertainty. 
The fraction of $B$ cascade decays in the $\Bbar \to X \ell^- \bar{\nu}_{\ell}$ sample is varied within its uncertainties and the differences in the $\Bbar \to X \ell^- \bar{\nu}_{\ell}$ signal yields are included in the systematic uncertainties. 
Possible differences in the $B_\mathrm{tag}$ composition of the MC simulation and data can affect the efficiencies and the cross-feed between charged and neutral $B$  events. 
To evaluate this effect we assume a conservative 30\% systematic uncertainty to the cross-feed fractions and we evaluate the systematic uncertainty by looking at differences in the measured values of ${\cal G}(1) |V_{cb}|$ and $\rho^2$ as we change the cross-feed fractions. 
We vary the $\Bbar \to D^{*} \ell^- \bar{\nu}_{\ell}$ form factors within their measured uncertainties~\cite{BABARFF} and we use an HQET parameterization~\cite{LLSW} for $\Bbar \to D^{**} \ell^- \bar{\nu}_{\ell}$. We also vary the $\Bbar \to D^{**}\ell^-\bar{\nu}_{\ell}$ branching fractions within the measured uncertainties~\cite{pdg}. For the $|V_{cb}|$ measurement, we include a contribution due to the uncertainties on the branching fractions of the reconstructed $D$  modes, and on the absolute branching fraction ${\cal B} (\Bbar \to X \ell^- \bar{\nu}_{\ell})$ used for the normalization.

\begin{table}[!th]
\centering
\caption{Systematic uncertainties in the measurement of ${\cal G}(1)|V_{cb}|$ and $\rho^2$  for $\overline{B} \rightarrow D \ell^- 
\bar{\nu}_{\ell}$ decays. We report the relative error (in \%)  for ${\cal G}(1)|V_{cb}|$ and the absolute error on $\rho^2$.}
\begin{tabular}{|l|c|c|c|c|c|c|}
\hline
\hline
 &\multicolumn{6}{c|}{{\scriptsize Systematic uncertainty on $|V_{cb}|$ and $\rho^2$ }} \\
\hline
& \multicolumn{2}{c|}{$D^0\ell^-\bar{\nu}_{\ell}$} & \multicolumn{2}{c|}{$D^{+}\ell^-\bar{\nu}_{\ell}$} & \multicolumn{2}{c|
}{$D\ell^-\bar{\nu}_{\ell}$} \\
\hline
& $|V_{cb}| (\%)$ & $\rho^2$ & $|V_{cb}| (\%)$ & $\rho^2$ & $|V_{cb}| (\%)$ & $\rho^2$ \\
\hline
Tracking efficiency      & 0.5 &0.008 & 1.1 & 0.003 & 0.7 & 0.004\\
Neutral reconstruction   & 1. &0.003 & 0.8 &0.006 & 0.9 &0.004\\
Lepton ID                & 1.0 & 0.009    & 0.9  & 0.009   & 0.95    &0.009   \\
PHOTOS  & 0.13 & 0.005 & 0.10 & 0.005 & 0.12 &  0.005\\
Cascade $\Bbar \to X \to \ell^-$ decay background  & 0.6 & - & 1.0 & - & 0.75 &  -\\
$\Bbar-B^-$ cross-feed & 0.24& 0.003 & 0.24 & 0.003 & 0.24 & 0.003 \\
$\Bbar \to D^{*}\ell^-\bar{\nu}_{\ell}$ Form factors &0.56 &0.008 & 0.20 & 0.003 & 0.38  & 0.006\\
$\Bbar \to D^{**}\ell^-\bar{\nu}_{\ell}$ Form factors & 0.24 &0.007& 0.34 & 0.006 & 0.29  &0.007\\
$D$ branching fractions & 1.0 &-& 1.35 & - & 1.12 &- \\
${\cal B}(\Bbar \to D^{**}\ell^-\bar{\nu}_{\ell})$ & 1.18 &0.023& 0.96 & 0.011 & 1.08  &0.019 \\
${\cal B} (\overline{B} \to X \ell^- \bar{\nu}_{\ell})$ & 0.95  & - & 0.95 & -  & 0.85 & -\\
$B_\mathrm{tag}$ selection & 1.1 & 0.021 & 1.8 & 0.036 & 1.5 & 0.028\\
$\Bbar \to X \ell^- \bar{\nu}_{\ell}$ yield  &0.7 & - & 1.1 & - & 0.85 & - \\
$\Bbar \rightarrow D \ell^- \bar{\nu}_{\ell}$ yield &1.27 & 0.018 & 1.06 & 0.027 & 1.25 & 0.020 \\
\hline
 Total systematic error & 3.1 & 0.04 & 3.6 & 0.05 & 3.3& 0.04\\
\hline
\hline
\end{tabular}
\label{tab:Syst1}
\end{table}

We also evaluate a systematic uncertainty due to differences in the efficiency of the $B_\mathrm{tag}$ selection in the exclusive selection of $\Bbar \to D \ell^- \bar{\nu}_{\ell}$ decays and the inclusive $\Bbar \to X \ell^- \bar{\nu}_{\ell}$ reconstruction, by using the same $B_\mathrm{tag}$ candidate selection adopted in the $\Bbar \to X \ell^- \bar{\nu}_{\ell}$ reconstruction also for the $\Bbar \to D \ell^- \bar{\nu}_{\ell}$ decays, and taking the difference in the signal yield, corrected for the difference in the  reconstruction efficiency, as a systematic uncertainty.

The systematic uncertainty in the determination of the $\Bbar \to X \ell^- \bar{\nu}_{\ell}$ yield is estimated by using an alternative fit method, which is then compared to the result of the nominal  $m_{ES}$ fit. We consider the $m_{ES}$ distribution from the data and the combinatorial \BB\, continuum and other background components (cascade and hadronic $B$ decays) modeled with distributions taken from the Monte Carlo simulation. We fit the background normalization on data in the $m_{ES}$ sideband region, defined by $m_{ES} < 5.265$ GeV/$c^2$. The normalization for the continuum background is fixed to the value obtained from off-peak data. The total background contribution is then subtracted from the total number of events in the $m_{ES}$ distribution to extract the  $\Bbar \to X \ell^- \bar{\nu}_{\ell}$ signal yield.  The uncertainty in the determination of the $\Bbar \to D\ell^- \bar{\nu}_{\ell}$ yield  in the different $w$ intervals is estimated by changing the PDFs used to model the different contributions in the $m^2_\mathrm{miss}$ distribution, e.g. by replacing the continuum PDFs with the corresponding one obtained from off-peak data.

\section{Results}
\label{sec:conclusions}

We present a measurement of ${\cal G}(1) |V_{cb}|$ and $\rho^2$ for $\Bbar \to D \ell^- \bar{\nu}_{\ell}$.  For $B^- \to D^0 \ell^- \bar{\nu}_{\ell}$, we obtain:

\begin{eqnarray}
{\cal G}(1)|V_{cb}|&=&(41.7 \pm 2.1 \pm 1.3 )\times 10^{-3} \nonumber \\
\rho^2 &=& 1.14 \pm 0.11 \pm 0.04 \nonumber \\ 
{\cal B} (B^- \to D^0 \ell^- \bar{\nu}_{\ell}) &=& (2.31 \pm 0.08 \pm 0.07)\%, \\ \nonumber
\end{eqnarray}

\noindent while for $\Bzb \to D^+ \ell^- \bar{\nu}_{\ell}$, we obtain:

\begin{eqnarray}
{\cal G}(1)|V_{cb}|&=&(45.6 \pm 3.3 \pm 1.6 )\times 10^{-3} \nonumber \\
\rho^2&=& 1.29 \pm 0.14 \pm 0.05 \nonumber \\
{\cal B} (\Bzb \to D^+ \ell^- \bar{\nu}_{\ell}) &=& (2.23 \pm 0.11 \pm 0.08)\%. \\ \nonumber
\end{eqnarray}

\noindent The results of the combined fit are:

\begin{eqnarray}
{\cal G}(1)|V_{cb}|&=&(43.0 \pm 1.9 \pm 1.4)\times 10^{-3} \nonumber \\
\rho^2&=& 1.20 \pm 0.09 \pm 0.04 \nonumber \\ 
{\cal B} (\Bzb \to D^+ \ell^- \bar{\nu}_{\ell}) &=& (2.17 \pm 0.06 \pm 0.07)\%. \\ \nonumber
\end{eqnarray}

\noindent 
Using an unquenched lattice calculation \cite{Okamoto}, corrected by a factor of 1.007 for QED effects, we get

\begin{eqnarray}
|V_{cb}|&=&(39.8\pm 1.8 \pm 1.3 \pm 0.9_{FF})\times 10^{-3} 
\end{eqnarray}

\noindent where the third error is due to the theoretical uncertainty in ${\cal G}(1)$. 
The resulting value of $|V_{cb}|$ is fully compatible with the other existing measurements on $\Bbar \to D\ell^-\bar{\nu}_{\ell}$,
and also with the measurement obtained using $\Bbar \to D^*\ell^-\bar{\nu}_{\ell}$.
Within the total errors this measurement is also compatible with the inclusive determination
of $|V_{cb}| = (41.68 \pm 0.39 \pm 0.58) \times 10^{-3}$~\cite{hfag}.

\section{Acknowledgments}
\label{sec:Acknowledgments}

We are grateful for the 
extraordinary contributions of our \pep2\ colleagues in
achieving the excellent luminosity and machine conditions
that have made this work possible.
The success of this project also relies critically on the 
expertise and dedication of the computing organizations that 
support \babar.
The collaborating institutions wish to thank 
SLAC for its support and the kind hospitality extended to them. 
This work is supported by the
US Department of Energy
and National Science Foundation, the
Natural Sciences and Engineering Research Council (Canada),
the Commissariat \`a l'Energie Atomique and
Institut National de Physique Nucl\'eaire et de Physique des Particules
(France), the
Bundesministerium f\"ur Bildung und Forschung and
Deutsche Forschungsgemeinschaft
(Germany), the
Istituto Nazionale di Fisica Nucleare (Italy),
the Foundation for Fundamental Research on Matter (The Netherlands),
the Research Council of Norway, the
Ministry of Education and Science of the Russian Federation, 
Ministerio de Educaci\'on y Ciencia (Spain), and the
Science and Technology Facilities Council (United Kingdom).
Individuals have received support from 
the Marie-Curie IEF program (European Union) and
the A. P. Sloan Foundation.


\begin{thebibliography}{99}
\bibitem{charge}
Charge conjugate states and transitions are always implied, unless explicitely stated otherwise.
\end{thebibliography}
\end{document}